\documentclass[12pt]{article}

\usepackage{color}
\usepackage[dvips]{graphicx,psfrag}
\usepackage{amsmath,amssymb}
\usepackage{bm}
\baselineskip=\normalbaselineskip
\renewcommand{\baselinestretch}{1.4}
\newcommand{\resection}[1]
 {\setcounter{equation}{0}\section{\large{#1}}}

\renewcommand{\thefootnote}{\fnsymbol{footnote}}
\newcommand{\bel}[1]{\begin{equation}\label{#1}}
\newcommand{\bal}[1]{\begin{eqnarray}\label{#1}}

\newcommand{\be}{\begin{equation}}
\newcommand{\ee}{\end{equation}}
\newcommand{\ba}{\begin{eqnarray}}
\newcommand{\ea}{\end{eqnarray}}

\newcommand{\eq}[1]{(\ref{#1})}

\begin{document}
\setcounter{page}{0}
\begin{flushright}
\parbox{40mm}{%
KUNS-1851 \\
YITP-03-39\\
{\tt hep-th/0307061} \\
July 2003}

\end{flushright}

\vfill

%%%%%%%%%%%%%%%%%%%%%%%%%%%%%%%%%%%%%%%%%%%%%%%%%%%%%
%% Title
%%%%%%%%%%%%%%%%%%%%%%%%%%%%%%%%%%%%%%%%%%%%%%%%%%%%
\begin{center}
{\large{\bf 
Limiting Temperature, Limiting Curvature \\
and the Cyclic Universe
}}
\end{center}

\vfill

%%%%%%%%%%%%%%%%%%%%%%%%%%%%%%
%% author
%%%%%%%%%%%%%%%%%%%%%%%%%%%%%%
\begin{center}
{\sc Masafumi Fukuma}$^{1)}$%
\footnote{E-mail: {\tt fukuma@gauge.scphys.kyoto-u.ac.jp}},  
{\sc Hikaru Kawai}$^{1)2)}$%
\footnote{E-mail: {\tt hkawai@gauge.scphys.kyoto-u.ac.jp}} and 
{\sc Masao Ninomiya}$^{3)}$%
\footnote{E-mail: {\tt ninomiya@yukawa.kyoto-u.ac.jp}} 
%\\[2em]

~\\

$^{1)}${\sl Department of Physics, Kyoto University, Kyoto 606-8502, Japan} \\
$^{2)}${\sl Theoretical Physics Laboratory, RIKEN, Wako 351-0198, Japan} \\
$^{3)}${\sl Yukawa Institute for Theoretical Physics, 
      Kyoto University, Kyoto 606-8502, Japan } \\

\end{center}

\vfill
%%%%%%%%%%%%%%%%%%%%%%%%%%%%%%%%%%
% Main
%%%%%%%%%%%%%%%%%%%%%%%%%%%%%%%%%%
\renewcommand{\thefootnote}{\arabic{footnote}}
\setcounter{footnote}{0}
\addtocounter{page}{1}
%%%%%%%%%%%%%%%%%%%%%%%%%%%%%%%%%%%%%%%%%%%%%%%%%%%%%%%%%%%%%%%%
%%%%%%%%%%%%%%%%%%%%%%%%%%%%%%%%%%%%%%%%%%%%%%%%%%%%%%%%%%%%%%%%
%%%%%%%%%%%%%%%%%%%
%% Abstract
%%%%%%%%%%%%%%%%%%%

\begin{center}
{\bf abstract}
\end{center}

\begin{quote}
\small{

We propose a new type of cosmological model 
in which it is postulated that not only the temperature but also the curvature 
is limited by the mass scale of the Hagedorn temperature. 
We find that the big bang of this universe is smoothly connected 
to the big crunch of the previous universe 
through a {\em Hagedorn universe}, 
in which the temperature and curvature remain 
very close to their limiting values. 
In this way, we obtain the picture of a cyclic universe. 
By estimating the entropy gained in each big crunch and big bang, 
we reach the conclusion that our universe has repeated 
this process about forty times 
after it was created at the Planck scale. 
We also show that the model gives a scale-invariant 
spectrum of curvature perturbations. 
}
\end{quote}
\vfill
%%
%\baselineskip=\normalbaselineskip
\renewcommand{\baselinestretch}{1.4}
%\setlength{\parskip}{0.3\baselineskip}
%%%%%%%%%%%%%%%%%%%%%%%%%%%%%%%%%%%%%%%%%%%%%%%%%%%%%%%%%%%%%%%%
%%%%%%%%%%%%%%%%%%%%%%%%%%%%%%%%%%%%%%%%%%%%%%%%%%%%%%%%%%%%%%%%
\newpage
%%%%%%%%%%%%%%%%%%%%%%%%%%%%%%%%%%%%%%%%%%%%%%%%%%%%%%%%%%%%%%%%
%%%%%%%%%%%%%%%%%%%%%%%%%%%%%%%%%%%%%%%%%%%%%%%%%%%%%%%%%%%%%%%%

\resection{Introduction}

There have been significant developments in observational cosmology 
in recent years.
In particular, many of the relevant cosmological parameters 
have been determined by data obtained recently from WMAP \cite{WMAP}, 
and a quantitative description of the structure of the very early universe 
is beginning to emerge. 
At present, analysis of the anisotropy of the cosmic
microwave background (CMB) seems to support the picture of an inflationary
universe \cite{inflation}\cite{Liddle}.
Although the idea of an inflationary universe resolves some
unnaturalness in the standard cosmology, 
the mechanism responsible for it is still to be clarified.
For instance, if we assume that the inflation is caused 
by the vacuum energy of a scalar field called an inflaton, 
we have to incorporate the potential, 
which would require a fine-tuning 
that is not natural in ordinary field theories.
Thus, the inflationary scenario removes unnaturalness in some ways and 
introduces it in others.
It might be thought that the era before the inflation should be
treated only after the construction of quantum gravity, 
but recent observational results  provide information 
regarding the time around this era. 
In this sense, cosmology is entering a new phase, in which 
we will be able to describe the very early universe,
 including the pre-inflation era.
In this respect, models of the cyclic universe 
deserve to be reconsidered as possible scenarios 
of the pre-big-bang universe. 

The basic idea of the cyclic universe is 
that the universe has repeated big bangs and big crunches many
times and has stored entropy through these processes, yielding 
the large amount that exists at present. 
This understanding of the universe was first proposed 
in the 1930s \cite{cyclic1}. 
Recently it has been argued \cite{cyclic2} that
this picture is not at all absurd from a physical point of view and, 
on the contrary, that it can be scientifically verified.

Our viewpoint in the present article is close to that of the cyclic universe.
However, we assume only the basic properties of string theory, 
and we seek to determine what we would see if we traced back 
from the present time to the big bang. 
We consider the closed Friedmann-Robertson-Walker (FRW) universe 
with radius $a(t)$ and rescale the comoving coordinates such that $k=1$. 

Before the matter-dominated era, our universe was dominated by radiation. 
During that radiation-dominated era, 
the radiation evolved adiabatically in time. 
As we see below, a field theoretical analysis shows 
that this adiabaticity no longer holds 
when the temperature becomes close to the Planck mass. 
The radius of the universe at such a time is estimated to be 
$10^{30}$ times the Planck length 
if we assume $\Omega_0=1.02$. 
In order to study the universe prior to that time, 
we need to take into account stringy effects that become important 
in two senses: 
Because of these effects, first, 
we should use an equation of state of the Hagedorn type 
for the radiation and, second, 
the Einstein equation that describes the time evolution of the space-time 
is subject to large corrections. 

In string theory, the temperature cannot exceed the upper bound, 
that is, the Hagedorn temperature $T_H$. 
There, the Einstein equation is no longer valid, 
because the energy density is much larger than the Planck scale, 
and higher excited states give rise to large corrections. 
Taking into account the fact that the expanding universe 
has the Hawking temperature proportional to the 
Hubble parameter $H=\dot a/a$ \cite{GH}, 
we can conclude that the  Hubble parameter also 
has an upper bound. 
If we assume that this is the case before the radiation-dominated era, 
we find that the universe expands exponentially 
at the rate $\dot a/a=T_H$, 
while the temperature remains close to the Hagedorn temperature. 
We call such a universe the {\em Hagedorn universe}. 
In contrast to the temperature and the Hubble parameter, 
the energy density of the universe has a strong time dependence, 
growing exponentially with $a^{-3}$.

As stated above, 
the universe had a radius that is about $10^{30}$ times larger 
than the Planck length at the beginning of the radiation-dominated era.
Therefore, if we go back in time farther by a period of about $70l_s$, 
the universe would shrink to the size of the string scale, $l_s$, 
which is the minimum length in string theory. 
We thus expect that the universe would rebound at this time 
and start expanding. 

Thus we obtain a picture of a cyclic universe in which the big bang 
and big crunch are connected through the Hagedorn universe. 
We can estimate the amount by which the entropy changes when the universe 
undergoes such a big crunch or big bang. 
A rough estimation shows that the entropy is increased by 10--400 times 
when the universe passes to the next cycle. 
As mentioned above, 
the present universe possesses an entropy $\sim 10^{90}$ times 
that of the string scale. 
This rough analysis would thus imply that the present cycle of the universe 
is between its 35th and 45th. 

According to the results of a recent observation, 
the present universe is in a stage of accelerating expansion, and hence 
no big crunch will occur in the future.
However, it can be shown 
that even if this is correct the previous universe actually could have 
had a big crunch if its total entropy was less than 
$1/90$ of that of the present universe for $\Omega_0=1.04$ 
or $1/250$ for $\Omega_0=1.02$.
(Here we have assumed $\Omega_{{\rm m}\,0}:\Omega_{\lambda\,0}=1:2$ 
\cite{WMAP}.)

The Hagedorn universe is similar to the inflationary universe 
in that they both expand exponentially. 
However, it is different in that 
it is in a high-temperature state filled with radiation of 
large energy density.
Therefore, it is not at all obvious 
whether we can obtain theoretically the scale-invariant spectrum 
of the curvature perturbation 
that can be directly observed in the anisotropy of the CMB. 
We show here that in fact it can be obtained from our model 
if we assume that the zero modes of the radiation fields 
have relaxation times on the order of 20 to 30 times longer 
than the string scale. 
The manner in which the scale invariance holds is rather different from that 
of ordinary inflationary models.

The present article is organized as follows. 
In Section 2, we show that the universe during 
the radiation-dominated era evolves adiabatically 
only if the temperature is lower than the Planck energy. 
In Section 3 we describe some properties of the universe 
of the Hagedorn temperature  
and introduce the concept of the limiting curvature. 
We then argue that the entropy of the universe is produced 
through a big crunch and the subsequent big bang. 
In Section 4 
we demonstrate that the scale-invariant spectrum of 
the curvature perturbation is realized also in the Hagedorn universe. 
Section 5 is devoted to a conclusion and comments. 

%%%%%%%%%%%%%%%%%%%%%%%%%%%%%%%%%%%%%%%%%%%%%%%%%%%%%%%%%%%%%%%%
%%%%%%%%%%%%%%%%%%%%%%%%%%%%%%%%%%%%%%%%%%%%%%%%%%%%%%%%%%%%%%%%
\resection{The condition that the radiation evolves adiabatically} 
%%%%%%%%        Contents starts here  %%%%%%%%%%%%%%

If we trace back the present universe, 
before the matter-dominated era, 
the universe was dominated by radiation. 
Going back farther through the radiation-dominated era, 
the temperature of the radiation approaches the Planck mass $m_p$. 
The time evolution of the radiation while the temperature is 
below such a value 
can be shown to be adiabatic in the manner described below.

The metric of the closed universe is given by 
\begin{align} 
 ds^2=dt^2-a(t)^2 \sum_{i,j=1,2,3}\gamma_{ij}(\vec{x})\,dx^i dx^j , 
\end{align}
where $x^i$ $(i=1,2,3)$ are the comoving coordinates 
and $\gamma_{ij}(\vec{x})$ is the metric of the unit three-sphere
with $k=1$. 
Then the time evolution of $a(t)$ is determined by the Hamiltonian constraint%
\footnote{
In the present article, we use units in which $\hbar=c=k_{\rm B}=1$, 
and set $G=1/m_p^2$ and $\lambda\equiv \Lambda/6$. 
}
\begin{align} 
 H_{\rm tot}\equiv a\left(\frac{1}{2}\,\dot{a}^2
 -\frac{1}{m^2_p}\frac{E}{a}-\lambda a^2+\frac{k}{2}\right)
 =0.
\end{align}
Here, the total energy of the universe 
is given by $E\equiv(4\pi/3)\rho\,a^3$, with $\rho$ the energy density. 
For the radiation-dominated universe, it is given by
\begin{align}
 E\propto T^4a^3,
\end{align}
with $T$ the temperature.
When the universe is in the early stage, 
we can ignore the third and fourth terms 
in the above expression of $H_{\rm tot}$, 
so that we obtain 
\begin{align}
 \frac{\dot{a}}{a}\sim\frac{T^2}{m_p}.
\end{align}
This quantity has dimensions of (time)$^{-1}$, 
and it is a measure of the speed of the expansion of the universe. 
On the other hand, if $T$ is the temperature of the radiation, 
the typical frequency, $\omega$, is given by 
\begin{align}
 \omega\sim T.
\end{align}
Inserting the above two equations into the condition 
that the time evolution of the radiation be adiabatic,
\begin{align}
 \omega\gg\frac{\dot{a}}{a},
\end{align}
we obtain
\begin{align}
 T\ll m_p.
\end{align}
We thus find that when the temperature is lower than the Planck mass, the
time evolution of the radiation is adiabatic.

%%%%%%%%%%%%%%%%%%%%%%%%%%%%%%%%%%%%%%%%%%%%%%%%%%%%%%%%%%%%%%%%
%%%%%%%%%%%%%%%%%%%%%%%%%%%%%%%%%%%%%%%%%%%%%%%%%%%%%%%%%%%%%%%%

\resection{The Hagedorn universe}

%%%%%%%%%%%%%%%%%%%%%%%%%%%%%%%%%%%%%%%%%%%%%%%%%%%%%%%%%%%%%%%%

\subsection{The universe at the Planck temperature}

As stated in the introduction, 
if we take the value $\Omega_0=1.02$, 
we find that the radius of our universe was 
\begin{align}
 a\sim10^{30}\times l_p
\end{align}
when the temperature was close to the Planck mass $m_p$. 
This enormous factor is related to what is called the ``flatness problem". 
The time at which the radius reaches this value from 0 
is naively calculated to be $t\sim l_p$ using the 
relation $a\sim t^{\frac{2}{3}}$ 
that holds during the radiation-dominated era.
However, as discussed in the following subsections, 
this method of estimation may need considerable modification 
in order to accurately describe the real history of the universe. 
%%%%%
%%%%%
%
\begin{figure}[htbp]
\begin{center}
%\rotatebox{270}{
\resizebox{!}{75mm}{
\includegraphics{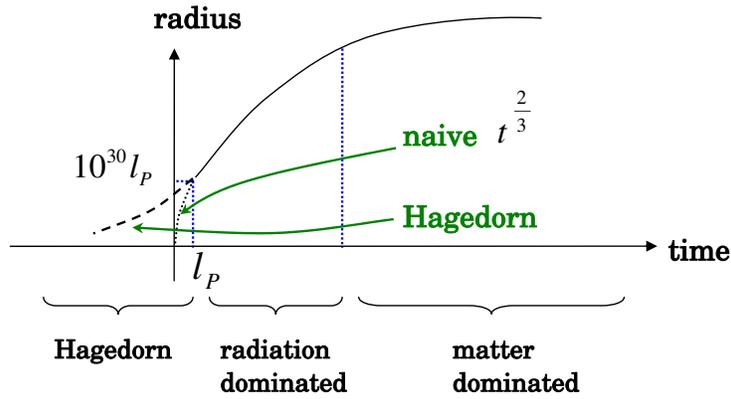}
%}
}
\end{center}
\caption{\footnotesize{Era of the Hagedorn universe 
existing before the radiation-dominated era.}}
\label{fig1}
\end{figure}
%
%%%%%
%%%%%

%%%%%%%%%%%%%%%%%%%%%%%%%%%%%%%%%%%%%%%%%%%%%%%%%%%%%%%%%%%%%%%%

\subsection{Preparation: upper bounds on the temperature 
and curvature}

As the temperature of the radiation approaches the Planck mass, 
stringy effects become important in two respects.
Firstly, the temperature has an upper bound, 
so that the proper equation of state becomes that of the Hagedorn type.
Secondly, there appears an upper bound also for the space-time curvature.
The first effect is well understood in string theory 
and is described in the introduction.%
\footnote{
The Hagedorn temperature $T_H$ is expressed in terms 
of the string mass scale $m_s$ as 
\begin{align}
 T_H=\left\{
\begin{array}{l}
m_s/2\sqrt{2}\pi\sim6\times 10^{16}\mbox{GeV (for the type II
strings),}\\
m_s/(2+\sqrt{2})\pi\sim 5\times 10^{16}\mbox{GeV (for the
heterotic strings).}
\end{array}
\right.
\end{align}
Strictly speaking, $m_s$ is approximately a factor of 10 smaller than $m_p$, 
and $T_H$ is also approximately a factor of 10 smaller than $m_s$, 
as can be seen from the above equation. 
However, hereafter we simply set $T_H$ equal to $m_s$, 
because the conclusion of this article
is not affected by this simplification.
}
We now explain the second effect. 

For any quantity obtained through differentiation of some physical quantity 
with respect to space-time coordinates, 
the expectation value should be bounded by the string mass scale. 
This is because string theory has a natural cutoff 
around the string mass $m_s$. 
Consideration of this fundamental point 
leads to the conclusion that the curvature is bounded by $m_s$: 
\begin{align}
 |R_{\mu\nu\lambda\rho}|<m_s^2.
\end{align}
A more convincing argument for the existance of such a bound can be given, 
at least for the 00-component of the Einstein tensor of the FRW universe, 
\begin{align}
 G_{00}=-3\frac{\dot{a}^2}{a^2}\sim-H^2,\quad
  \Bigl(H=\frac{\dot a}{a}\Bigr)
\end{align}
in the following way. 

We first note that the Hawking temperature is approximately 
equal to the Hubble parameter $H$. 
We prove this now for a massless scalar field, as a simple example.
In the FRW universe, 
the Hamiltonian of a massless scalar field is given by
\begin{align}
 H_{\rm matter}=\sum_{\vec{k}}
  \left(\frac{|\pi_{\vec{k}}|^2}{2a^3}
 +\frac{a^3\omega_{\vec{k}}^2 |\phi_{\vec{k}}|^2}{2}\right)
 =\sum_{\vec{k}}\left(\frac{|\pi_{\vec{k}}|^2}{2a^3}
  +\frac{a\vec{k}^2|\phi_{\vec{k}}|^2}{2}\right).
 \label{H_matter}
\end{align}
Here $\vec{k}$ represents a comoving wave number, which 
has the following dispersion relation with physical frequency 
$\omega_{\vec{k}}$: 
\begin{align}
 \omega_{\vec{k}}(t)=\frac{|\vec{k}|}{a(t)}.
 \label{one}
\end{align}

According to Eq.\ (\ref{one}), 
the frequency of the mode $\phi_{\vec{k}}$ becomes smaller 
as the universe expands.
To consider the Hawking radiation, 
we can use the following
approximation. 
As long as the physical frequency is larger than the Hubble parameter $H$, 
$\phi_{\vec{k}}$ evolves adiabatically in time.
However, when $\omega_{\vec{k}}$ becomes smaller than
$H$, $\phi_{\vec{k}}$ no longer changes in time, remaining 
at the value it took at the time $t_k$ 
when $\omega_{\vec{k}}$ became equal to $H$.
Denoting by $a_{\vec{k}}$ the radius of the universe at that moment, 
$a_{\vec{k}}=a(t_{\vec{k}})$, 
we obtain the equation
\begin{align}
\frac{a_{\vec{k}}}{a(t)}=\frac{\omega_{\vec{k}}(t)}{H}, 
\label{three}
\end{align}
because $\omega_{\vec{k}}(t_{\vec{k}})=H$. 

First, we assume that the scalar field state is in the ground state. 
Then, we can suppose that at the moment $t_{\vec{k}}$ 
when the field leaves the Hubble horizon, 
the mode $\phi_{\vec{k}}$ possesses the zero-point energy  
\begin{align}
 \frac{1}{2}\omega_{\vec{k}}\left(t_{\vec{k}}\right)=\frac{1}{2}H.
\end{align}
By taking the potential part of the Hamiltonian equal to half 
of the zero-point energy, 
we obtain 
\begin{align}
\frac{a_{\vec{k}}\vec{k}^2|\phi_{\vec{k}}|^2}{2}=\frac{1}{4}H.
\label{four}
\end{align}
Thus, the energy of the mode $\phi_{\vec{k}}$ at this moment $t_{\vec{k}}$ 
is given by
\begin{align}
 (\mbox{the energy of }\phi_{\vec{k}})\sim
  \frac{a\vec{k}^2|\phi_{\vec{k}}|^2}{2}
 =\frac{a}{a_{\vec{k}}}\frac{a_{\vec{k}}\vec{k}^2|\phi_{\vec{k}}|^2}{2}
 =\frac{H^2}{4\omega_{\vec{k}}}.
\end{align}
In the derivation of the last equation here, we ignored the kinetic
energy of the mode $\phi_{\vec{k}}$ and used the equations (\ref{three})
and (\ref{four}).  
It may be better to carry out a more detailed, quantum 
mechanical calculation. 
If this were done, the final result might change by a factor of two or three.
In summary, in our approximation, the energy of the Hawking radiation
of the de Sitter universe is given by
\begin{align}
 \frac{H^2}{4\omega}\theta\left(H-\omega\right).
\end{align}
Note that the energies of the lower frequency modes become much larger
than the naively estimated zero-point energy, $\frac{1}{2}\omega$.
In this sense, the radiation we consider is nontrivial.

Next, note that if the universe is in equilibrium at 
temperature $T$, the mode $\phi_{\vec{k}}$ must possess an energy $T$, 
according to the law of equipartition.
A more rigorous argument can be made using the Planck distribution, 
but, because we consider the region where the frequency of
$\phi_{\vec{k}}$ is smaller than $H$, the result will not differ greatly 
if $T\sim H$.

In the same manner as above, 
when we set the potential part of the Hamiltonian equal to 
the energy per mode assuming an equipartition, we obtain
\begin{align}
\frac{a_{\vec{k}}\vec{k}^2|\phi_{\vec{k}}|^2}{2}=\frac{1}{2}T.
\label{five}
\end{align}
Comparing (\ref{four}) and (\ref{five}), we find that the
energies of the lower frequency modes
have the same form in the cases of both Hawking radiation and 
black-body radiation with temperature $H/2$.
This implies that the de Sitter universe 
with the Hubble parameter $H$ has temperature $H/2$.%
\footnote{
A more precise analysis yields the Hawking temperature given by 
$T=H/2\pi$ \cite{GH}.
}

If we take stringy excitations into account, 
there is an upper bound on the temperature, 
the Hagedorn temperature, $T_H$ .
Combining this fact with the above result, 
we conclude that the Hubble parameter $H$ possesses
a similar upper bound.
Specifically, we can state that in the FRW universe, 
the Hubble parameter $\dot{a}/a$ is bounded by
$T_H/2$.
Although this observation is sufficient for the following argument,
we further deduce, using the fact that
$G_{00}=-3\left(\dot{a}/a\right)^2$, that the curvature itself 
is bounded
by $m_s^2$, as mentioned in the beginning of this section.%
\footnote{
One possibility for a Lagrangian that possesses such property is 
 ${\cal L}=\sqrt{-\det\left(g_{\mu\nu}-\kappa R_{\mu\nu}\right)}
 -\sqrt{-\det\left(g_{\mu\nu}\right)} \ .$
}

%%%%%%%%%%%%%%%%%%%%%%%%%%%%%%%%%%%%%%%%%%%%%%%%%%%%%%%%%%%%%%%%

\subsection{Tracing the universe back farther}

We now trace back the history of the universe farther, starting from 
the time of the Planck temperature, considered in Subsection 3.1. 
For the universe with this temperature, it is natural to assume that the 
Hubble parameter of the universe takes its limiting value, 
\begin{align}
 \frac{\dot{a}}{a}\sim m_s.
\end{align}
In this case, the radiation fields are in the Hagedorn state, 
with the temperature fixed to the Hagedorn temperature: 
\begin{align}
 T\sim m_s.
\end{align}
On the other hand, as we trace back the universe, the energy density will
exponentially increase from a value of order $m^4_s$.
In the Hagedorn state, the energy is given by
\begin{align}
 \frac{E}{V}=\left(T_H-T\right)^{-\alpha},
\end{align}
with $\alpha$ a constant.
Therefore, if the temperature is close to $T_H$, the energy of the 
radiation will mainly be carried by massive states, whose frequencies
are much larger than $m_s$, so that the condition of adiabaticity,
\begin{align}
 \omega\gg\frac{\dot{a}}{a},
\end{align}
will be satisfied.

In summary, we find that before the radiation-dominated era, 
the matter fields are in the Hagedorn state, 
while the metric evolves as in a de Sitter space, 
with the Hubble parameter fixed to a value of order $m_s$.
Here, the temperature of the matter fields is equal to the Hawking 
temperature of the universe, and it evolves adiabatically, as a
kind of extreme state.
We call this extreme universe the Hagedorn Universe.

As argued in Subsection 3.1, at the time of the transition 
from the Hagedorn universe to the radiation-dominated era, 
the radius of the universe is
$10^{30}$ times larger than the string length $l_s$.
However, if we trace the universe back farther, the radius
decreases exponentially, as $\sim \exp(m_st)$.
Therefore, when we trace back for a period of about $70l_s$, 
the radius of the universe becomes comparable to $l_s$.

We may naively think the radius will eventually 
become smaller than $l_s$.
However considering that 
in string theory, $l_s$ is the minimum length, 
or that it possesses T-duality as a symmetry, 
it is natural to conjecture 
that the radius of the universe actually starts 
to increase after it reaches the minimum value $l_s$.
More precisely, if we assume the T-duality
\begin{align}
 a\leftrightarrow\frac{l^2_s}{a},
\end{align}
the radius is given by
\begin{align}
 a(t)=\frac{l^2_s}{l_s e^{m_st}}=l_se^{-m_st}
\end{align}
for negative $t$.
%%%%%
%%%%%
%
\begin{figure}[htbp]
\begin{center}
%\rotatebox{270}{
\resizebox{!}{55mm}{
\includegraphics{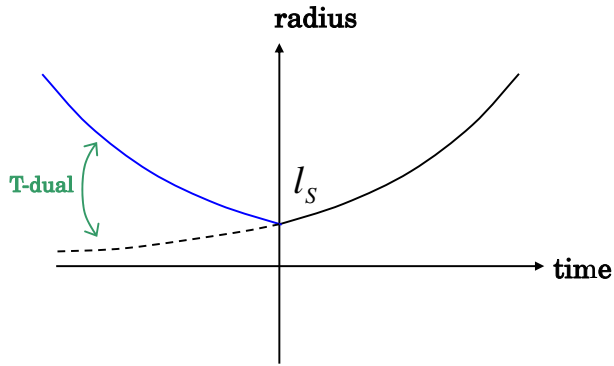}
}
%}
\end{center}
\caption{\footnotesize{The bounce of the universe due to T-duality.}}
\label{fig2}
\end{figure}
%
%%%%%
%%%%%
In other words, the time dependence of the radius $a$ is symmetric 
about its minimum (see Fig.\ \ref{fig2}).
Instead of using T-duality, we may hypothesize sipmly from the existence 
of the minimum length in string theory 
that the universe rebounds and begins to grow 
after its radius reaches the minimum value, $l_s$.

In the above argument, we have implicitly assumed that the topology of
the space is that of a three-torus. In fact, the argument becomes simpler 
if it is three-sphere. In this case, if the FRW metric has
the limiting curvature $l_s^{-2}$, 
it is identically that of the de Sitter space, 
\begin{align}
 ds^2=dt^2-a(t)^2d\Omega _3^2,\quad 
 a(t)=l_s \cosh(t),
\end{align}
which again seems to indicate that the universe rebounds and grows 
after it reaches the smallest size, $l_s$.

\begin{figure}[htbp]
\begin{center}
%\rotatebox{270}{
\resizebox{!}{60mm}{
\includegraphics{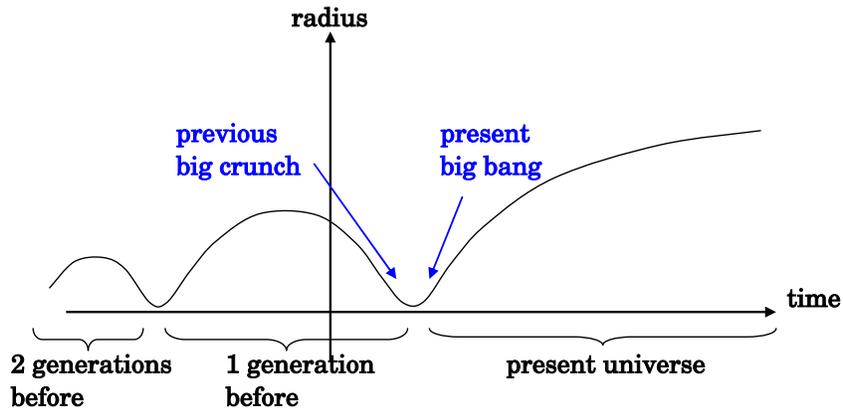}
}
%}
\end{center}
\caption{\footnotesize{ A sequence of cycles of the universe. }}
\label{fig3}
\end{figure}

At any rate, if we trace back the big bang, we see that the size of
the universe decreases to the order of the string length $l_s$,
at which point it rebounds and increases.
This implies that the big bang of our universe took place 
after the big crunch of the previous universe.
Thus we are lead to the picture of a cyclic universe 
(see Fig.\ \ref{fig3}).
We emphasize that this picture emerges rather naturally  
from simple assumptions, such as 
the existence of upper bounds on the temperature and curvature.
%%%%%
%%%%%
%
%
%%%%%
%%%%%

%%%%%%%%%%%%%%%%%%%%%%%%%%%%%%%%%%%%%%%%%%%%%%%%%%%%%%%%%%%%%%%%
%%%%%%%%%%%%%%%%%%%%%%%%%%%%%%%%%%%%%%%%%%%%%%%%%%%%%%%%%%%%%%%%

\subsection{Entropy production through big crunches and big bangs}

In this subsection we estimate the entropy 
produced by each big crunch and big bang. 
As we describe below, through each big crunch and big bang, 
the total entropy of the universe seems to become about 10--20 times 
larger than the previous value. 
If this is indeed the case, during a big crunch and the successive big bang, 
the total entropy will be increased by a factor of about 100--400. 
The present universe has an entropy that is about $10^{90}$ times larger than 
that of the naive Planck scale universe. In the picture presented here, 
this is interpreted 
as implying that the present universe has grown from the Planck size 
through 35--45 cycles. 

\begin{figure}[htbp]
\begin{center}
%\rotatebox{270}{
\resizebox{!}{40mm}{
\includegraphics{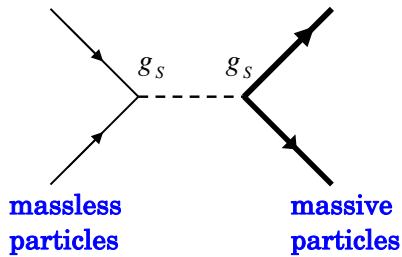}
}
%}
\end{center}
\caption{\footnotesize{ Massless particles 
converted to massive particles. }}
\label{fig4}
\end{figure}

Because the arguments regarding the big crunch and big bang are 
almost the same, 
we describe the calculation in detail for the case of the big crunch 
only in the rest of this subsection.
Entropy production occurs mainly in the transition period 
from the radiation-dominated era to the Hagedorn universe.
As the shrinking of the universe proceeds, 
the temperature reaches the string mass scale, 
and the process is subsequently no longer adiabatic. 
As described in Section 2, 
after this point a free massless field no longer evolves in time. 
The reason that the above-mentioned state passes into the Hagedorn state 
is that massless particles are converted to massive particles, 
due to the coupling between modes, that is, scattering 
(see Fig.\ \ref{fig4}).

\begin{figure}[htbp]
\begin{center}
%\rotatebox{270}{
\resizebox{!}{70mm}{
\includegraphics{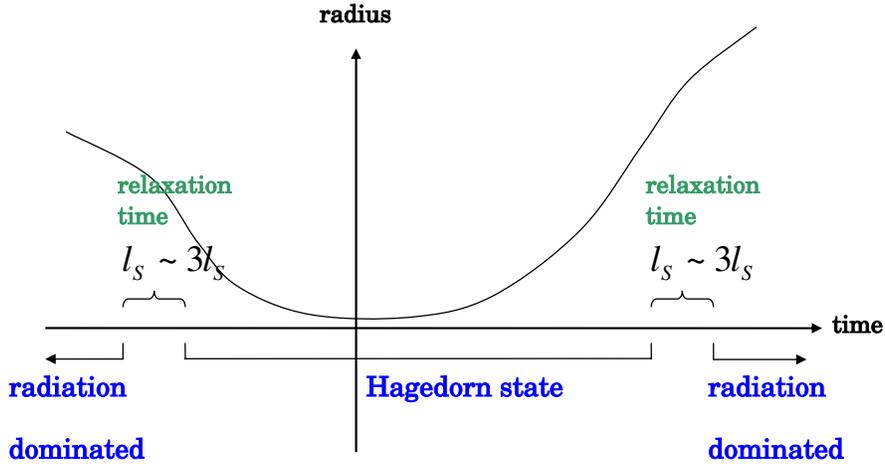}
}
%}
\end{center}
\caption{\footnotesize{ Transition of the radiation-dominated universe 
from one generation to the next, 
with the Hagedorn universe lying between. }}
\label{fig5}
\end{figure}

Let us estimate the relaxation time, which represents the time needed 
to produce massive particles through collisions of massless particles. 
When the temperature reaches $m_s$ in the radiation-dominated state, 
we can assume all the quantities 
appearing in the presently considered processes 
to be at the string scale. 
In this case, the relaxation time $\tau$ can be assumed to satisfy 
\begin{eqnarray}
        \tau \sim g_s^{-4}l_s.
 \label{3.20}
\end{eqnarray}
Here, $g_s$ is the string coupling constant, 
and therefore it may be thought from this equation that the relaxation time 
becomes approximately $10^2$--$10^3$ times larger than the string scale. 
However, because the Hagedorn universe shrinks exponentially in time, 
if, for example, during a time of $2l_s$, 
the radius of the universe would shrink by a factor of $\frac{1}{7.4}$. 
This implies an increase of the particle number density 
by a factor of 400, and therefore a decrease of the relaxation time 
by a factor of $\frac{1}{400}$. 
During a time of $3l_s$, 
the radius would change by a factor of $\frac{1}{20}$, 
and thus the relaxation time would change by a factor of $\frac{1}{8000}$. 
It is thus seen that even if the effect of the string coupling constant 
on Eq.\eq{3.20} is appreciable, the effect of the particle number density 
overwhelms this string coupling effect. 
For this reason, we can conclude that the time needed 
to pass from the radiation-dominated state into the Hagedorn state is 
of the order of $l_s$--$ 3l_s$ (see Fig.\ \ref{fig5}). 

We next estimate the amount of entropy produced during this time. 
To simplify the problem, we consider the following situation. 
Suppose that for times $t<0$, a massless field is 
in equilibrium at temperature $m_s$. 
We then change the size of the box containing these massless particles 
at a rate proportional to $e^{-m_st}$. 
As a result, these fields are no longer in equilibrium. 
We wish to determine how large the entropy is 
at a time $t=\tau$.

To solve this problem, we use the same Hamiltonian given in Eq.\ 
\eq{H_matter} in the form 
\begin{align}
 H_{\rm matter}&=\sum_{\vec{k}} \left( \frac{|\pi_{\vec{k}}|^2}{2a^3}
   +\frac{a\vec{k}^2|\phi_{\vec{k}}|^2}{2} \right) \\
 &=\sum_{\vec{k}} \omega_{\vec{k}}
   \left( \frac{|\pi_{\vec{k}}|^2}{2a^2|\vec{k}|}
   +\frac{a^2|\vec{k}|\phi_{\vec{k}}^2}{2} \right),
\end{align}
with
\begin{align}
 \omega_{\vec{k}}(t)=\frac{|\vec{k}|}{a(t)}.
\end{align} 
We further simplify the problem by assuming 
that the frequencies are cut off at $m_s$. 
In this case, we may use as a rather good approximation 
the law of equipartition, instead of the Planck distribution, 
in thermodynamic equilibrium. Then, the contribution to the Hamiltonian 
from the mode of frequency $\omega$ is expressed 
in terms of appropriate canonical coordinates $p$ and $q$ as 
\begin{align}
 H_{\omega}=\omega \left( \frac{1}{2}p^2+\frac{1}{2}q^2\right).
\end{align} 
The energy of this mode is equal to the temperature $m_s$, 
in accordance with the law of equipartition. 
The entropy can be estimated as follows. 
We consider the area in phase space defined by the relation 
\begin{align}
  \left(\frac{1}{2}p^2+\frac{1}{2}q^2\right)< m_s.
\end{align} 
The entropy we seek is obtained 
by taking the logarithm of this area 
after dividing by $2\pi$. This yields the value 
$\log \left( \frac{m_s}{\omega}\right)$.
Note that the frequency is cut off at $m_s$. 
Then, taking the average of the entropy over all modes, 
we obtain 
\begin{align}
        S_{\textit{before}}&=
        \frac{\displaystyle \int_{k\leq m_s}\frac{d^3k}{(2\pi)^3}
        \log\left(\frac{m_s}{k}\right)}{\displaystyle \int_{k\leq m_s}
        \frac{d^3k}{(2\pi)^3}\> 1}=\frac{1}{3}.
 \label{S_before}
\end{align}

\begin{figure}[htbp]
\begin{center}
%\rotatebox{270}{
\resizebox{!}{65mm}{
\includegraphics{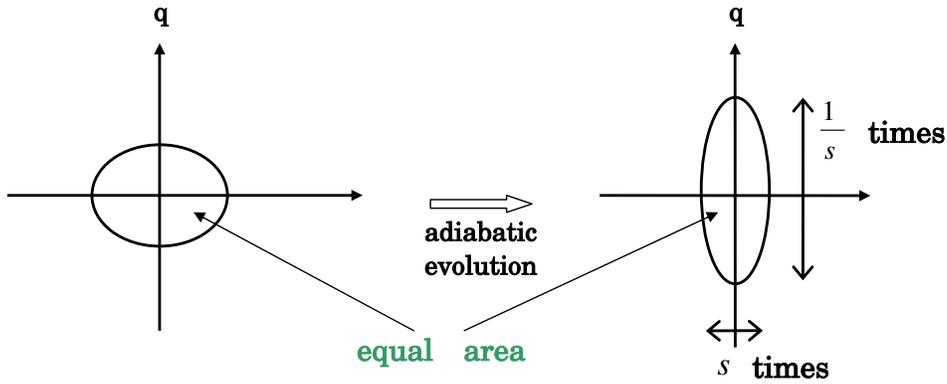}
}
%}
\end{center}
\caption{\footnotesize{ Adiabatic evolution: 
each orbit in the phase space has an area that remains fixed 
during the evolution.  }}
\label{fig6}
\end{figure}

We next estimate the amount of entropy produced from each mode 
when the radius of the universe changes. 
As in Subsection 3.2, 
we assume that modes with frequency less than $m_s$ 
are fixed at their initial values, 
because these modes evolve very slowly compared to 
the radius of the universe, 
which behaves as $a\propto e^{-m_st}$. 
The size of the universe shrinks by a factor of $s=e^{-m_s\tau}$ 
when the time passes from $t=0$ to $t=\tau$. 
If such a change occurred very slowly compared to the oscillation, 
each mode would evolve adiabatically in time. 
In phase space, 
the elliptic orbit representing the oscillation would change shape 
in such a manner that its area is constant. 
The equi-energy surface which is given by the equal area 
in the phase space changes 
by such that the $p$- and $q$-axes are enlarged 
$s$ times and $s^{-1}$ times, respectively (see Fig.\ \ref{fig6}).

\begin{figure}[htbp]
\begin{center}
%\rotatebox{270}{
\resizebox{!}{70mm}{
\includegraphics{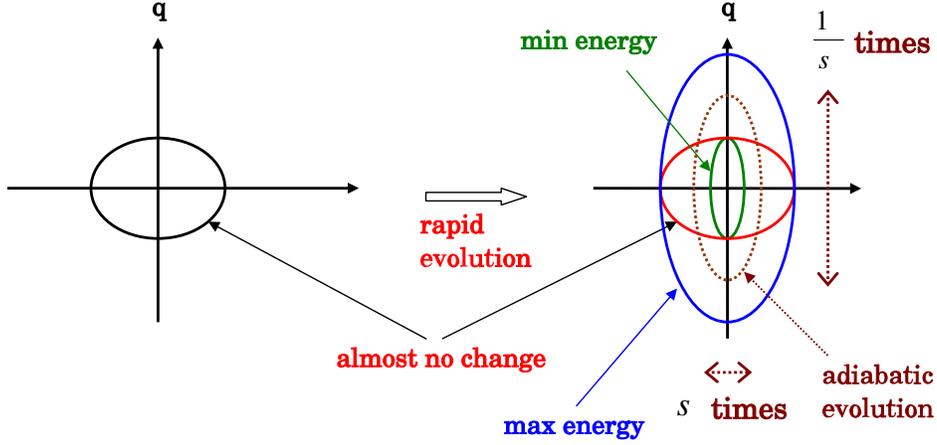}
}
%}
\end{center}
\caption{\footnotesize{ Non-adiabatic process due 
to the rapid evolution of the universe: 
each orbit is enlarged asymmetrically during such evolution, 
leading to the generation of entropy. }}
\label{fig7}
\end{figure}

However, in reality the radius of the universe is relieved 
to change very rapidly, so that the modes would evolve only very 
negligibly during the time that the radius changes by a factor of $s$. 
Therefore, if we consider an ensemble of points 
distributed uniformly on an equi-energy surface 
before the universe shrinks,
we can regard them as being distributed on the same surface 
even after the universe has undergone the rapid shrinking.  
On the other hand, as described above, 
if the change were adiabatic, 
an equi-energy surface would take the form of an ellipse, 
as in Fig.\ \ref{fig6}.  
Thus the energy of this distribution fluctuates from a value $s^2$ times 
to a value $s^{-2}$ times that in the case of adiabatic evolution. 
When the energy takes the maximal value, the area enclosed 
by the corresponding equi-energy surface 
is $s^{-2}$ times the initial area. 
This case is depicted in Fig.\ \ref{fig7}. 
\noindent
Thus, the increase of the entropy of this mode during this process 
is $\log s^{-2}$. 
In the present case, using $s=e^{-m_s\tau}$, we can conclude 
the entropy production per mode that occurs during the big crunch 
is $2m_s\tau$. 
Therefore, considering Eq.\ \eq{S_before}, 
the entropy per mode after the big crunch is given by 
\begin{align}
 \displaystyle \frac{\> S_{\textit{after}}\>}
{\> S_{\textit{before}}\>}=\frac{\> 
 \frac{1}{3}+2m_s\tau\>}{\frac{1}{3}}=1+6m_s\tau.
\end{align} 
To obtain the numerical value of this ratio, we set $\tau =l_s\,$--$\,3l_s$, 
and finally obtain 
\begin{align}
 \displaystyle \frac{\> S_{\textit{after}}\>}
 {\> S_{\textit{before}}\>}=7\,{\rm -}\, 20,
 \label{S_ratio}
\end{align} 
which supports the argument given in the beginning of this section.

Finally, we note that the value given in Eq.\ \eq{S_ratio} implies 
that the universe one generation before the present 
indeed ended in a big crunch. 
According to a recent observation, the present universe is already 
in the stage of accelerating expansion, 
and a big crunch will not occur in this generation. 
However, it can be shown that the previous universe actually could have 
had a big crunch if its total entropy was less than 
$1/90$ times that of the present universe for $\Omega_0=1.04$, 
or $1/250$ times for $\Omega_0=1.02$.
(Here, we have assumed $\Omega_{{\rm m}\,0}:\Omega_{\lambda\,0}=1:2$ 
\cite{WMAP}.) 
By taking into account the contribution 
from both the big crunch and the big bang, 
we find that for this to be the case, 
it is sufficient that the square of the value 
in Eq.\ \eq{S_ratio} be larger than $\sim$ 90--250.

It is thought that in previous generations 
the entropies were smaller, 
and we are thus led to conclude that the universe has grown 
through repeated big crunches and big bangs. 
Our universe represents the final stage of these cyclic growth, 
and it is now in the era of accelerating expansion. 
This conclusion may seem to assign an unnaturally special position 
to our universe. 
However, we can give an explanation 
akin to the anthropic principle 
that justifies the ``specialness" of the present universe: 
In each previous universe, 
the interval between the big crunch 
and the big bang was too short 
for life to emerge.

%%%%%%%%%%%%%%%%%%%%%%%%%%%%%%%%%%%%%%%%%%%%%%%%%%%%%%%%%%%%%%%%
%%%%%%%%%%%%%%%%%%%%%%%%%%%%%%%%%%%%%%%%%%%%%%%%%%%%%%%%%%%%%%%%

\resection{Cosmological perturbation of the Hagedorn universe}

In this section, we demonstrate that our model of the Hagedorn universe 
gives rise to a scale-invariant spectrum of curvature perturbations 
on the superhorizon scale.  
We here assume that thermal fluctuations are adiabatic 
and have an isotropic stress tensor.

The standard procedure to obtain the curvature fluctuations 
in the superhorizon is as follows \cite{perturbation}. 
We first calculate the velocity field $v(\vec{x},t)$ 
in the primordial era from the energy-momentum tensor 
using the relation 
\begin{align}
 T_{0i}&\equiv a\,\bigl(\bar{\rho}+\bar{p}\bigr)\,\partial_i v.
\end{align}
Here $\bar{\rho}(t)$ and $\bar{p}(t)$ are the background values 
of energy density and pressure, respectively. 
We also compute the potential perturbation $\Psi$.%
\footnote{
%%%%
Here, the metric under scalar perturbations is parametrized  
in the longitudinal gauge: 
\begin{align}
 ds^2=\bigl(1+2\,\Psi(\vec{x},t)\bigr)\,dt^2
  -\bigl(1+2\,\Phi(\vec{x},t)\bigr)\,a^2(t)\,
  \gamma_{ij}(\vec{x})\,dx^i\,dx^j,
\end{align}
where $\gamma_{ij}(\vec{x})\,dx^i\,dx^j$ is the metric of 
the unit three-sphere. 
The relation $\Psi+\Phi=0$ holds 
when the anisotropic stress tensor vanishes. 
If it is further assumed that the perturbations are adiabatic, 
then the combination $\Psi(\vec{x},t)+H(t)\, a(t)\,v(\vec{x},t)
\,\bigl[\equiv\zeta(\vec{x})\bigr]\,$ gives a constant of motion 
in the superhorizon, 
as mentioned in the main text. 
%%%%
}
Knowing the values of $v(\vec{x},t)$ and $\Psi(\vec{x},t)$ 
in the primordial era 
is sufficient to determine their values at later times.  
In fact, $\zeta(\vec{x})\equiv \Psi(\vec{x},t)+H(t)\, a(t)\, v(\vec{x},t)$ 
is independent of time in the superhorizon, 
and it can easily be shown from the Einstein equation that 
$\Psi=(2/3)\zeta$ for the radiation-dominated era 
and $\Psi=(3/5)\zeta$ for the matter-dominated era.

We now recall that our model is based on the assumption 
that radiation in the Hagedorn universe is in thermodynamical equilibrium 
with the constant Hagedorn temperature $T_H\sim m_s$. 
We also assume that each time slice has the topology of a three-sphere, 
and that its radius  $a(t)$ 
shrinks during the last stage of the former cycle 
and then re-expands in the beginning of the present cycle, 
in both cases evolving as the exponential $a(t)=l_s\,e^{\,m_s\,|t|}$.

Some massless modes in the previous cycle, 
$(\phi_{\vec{k}},\pi_{\vec{k}})$, 
enter the subhorizon region $\bigl(a(t)/k<H^{-1}\sim m_s\bigr)$ 
as the universe shrinks exponentially, 
and have a chance to evolve adiabatically there. 
The time $t_k$ at which the mode of wave number $\vec{k}$ 
exits the subhorizon region 
and reenters  the superhorizon region can be evaluated 
by setting the physical wavelength equal to the Hubble distance: 
$a(t_k)/k\sim H^{-1}\,\bigl(\sim l_s=1/m_s\bigr)$. 
A simple calculation gives $t_k=m_s \log k$, 
at which the universe has a size $a(t_k)\sim k\,l_s=k/m_s$.

{}Firstly, the amplitude of the velocity field in the Hagedorn universe 
can be calculated in the following way. 
Because the energy-momentum tensor of a massless field $\phi$ is given by 
$T_{\mu\nu}=\partial_\mu\phi\,\partial_\nu\phi
 -(1/2)\,(\nabla\phi)^2\,g_{\mu\nu}$, 
the Fourier component of the velocity field $v$ is given by
\begin{align}
 H a v_{\vec{k}}&=H\sum_{k'}
  \frac{ \dot{\phi}_{\vec{k}'}\,\phi_{\vec{k}-\vec{k'}}}{\bar{\rho}+\bar{p}}
  \nonumber\\
 &=H\sum_{k'}\frac{\pi_{\vec{k}'}\,\phi_{\vec{k}-\vec{k}'}}
  {a^3\,\bigl(\bar{\rho}+\bar{p}\bigr)}\nonumber\\
 &=\frac{H}{M}\,\sum_{\vec{k}'}\pi_{\vec{k}'}\,\phi_{\vec{k}-\vec{k}'}. 
\end{align} 
Here, in deriving the second line on the right-hand side, 
we have used the relation $\pi_{\vec{k}}(t)=a^3(t)\,\dot{\phi}(t)$. 
{}For the third line, we have assumed that 
$M\equiv a^3(t)\,\bigl(\bar{\rho}(t)+\bar{p}(t)\bigr)$ 
is a constant in time, 
because the energy density in the Hagedorn universe  
consists mainly of contributions from massive excitations of strings, 
so that it should evolve as in the matter-dominated era: 
\begin{align}
 a^3(t)\,\bigl(\bar{\rho}(t)+\bar{p}(t)\bigr)\sim
  a^3(t)\,\bar{\rho}(t) \equiv M\quad \bigl(\mbox{independent~of~$t$}\bigr).
\end{align}
The power spectrum of $H a(t) v(\vec{x},t)$ can thus be calculated 
by assuming the translational invariance of thermal fluctuations, 
and we obtain
\begin{align}
 \Bigl\langle\,\bigl|\,H a v_{\vec{k}}\bigr|^2\Bigr\rangle(t)
  &=\Bigl\langle\,\,H a v_{\!-\vec{k}}\,\, H a v_{\vec{k}}
   \Bigr\rangle(t) \nonumber\\
&=\biggl(\frac{H}{M}\biggr)^2\,\sum_{\vec{k}'}
  \Bigl\langle\,\bigl|\,\pi_{\vec{k}'}\bigr|^2\Bigr\rangle(t)\,
  \Bigl\langle\,\bigl|\,\phi_{\vec{k}-\vec{k}'}\bigr|^2\Bigr\rangle(t).
\end{align}

We note here that if the relaxation time $\tau(k')$ of a mode 
$(\phi_{\vec{k}'},\pi_{\vec{k}'})$ 
is longer than the duration of the Hagedorn universe, 
then such mode cannot fluctuate thermodynamically in the subhorizon region 
and takes a constant value during this era. 
Denoting by $k_c$ the maximum wave number of such frozen modes,%
%%%%
\footnote{
That is, $k_c$ is defined by the following equation:
\begin{equation}
 \tau(k_c)=\bigl(\mbox{period of time of the Hagedorn universe}\bigr)
  \sim 100\,l_s.
\end{equation}
}
%%%%
we separate the summation into two parts:
\begin{align}
 &\Bigl\langle\,\bigl|\,H a v_{\vec{k}}\bigr|^2\Bigr\rangle(t)\nonumber\\
 &= \biggl(\frac{H}{M}\biggr)^2 
  \left[\, \sum_{k'<k_c}
  \Bigl\langle\,\bigl|\,\pi_{\vec{k}'}\bigr|^2\Bigr\rangle(t)\,
  \Bigl\langle\,\bigl|\,\phi_{\vec{k}-\vec{k}'}\bigr|^2\Bigr\rangle(t)
  + \sum_{k'>k_c}
  \Bigl\langle\,\bigl|\,\pi_{\vec{k}'}\bigr|^2\Bigr\rangle(t)\,
  \Bigl\langle\,\bigl|\,\phi_{\vec{k}-\vec{k}'}\bigr|^2\Bigr\rangle(t)
  \,\right].
 \label{separate}
\end{align}
{}For the first term on the right-hand side, 
the mode $\phi_{\vec{k}-\vec{k}'}(t)$ is in the subhorizon for $t<t_k$ 
and fluctuates thermodynamically. 
Because the amplitude varies smoothly as a function of $\vec{k}'$, 
and because $k_c$ is very small, 
the amplitude is approximated well by replacing $\phi_{\vec{k}-\vec{k}'}(t)$ 
with $\phi_{\vec{k}}(t)$.
Then, again using the law of equipartition, we obtain
\begin{align}
 \Bigl\langle\,\bigl|\,\phi_{\vec{k}-\vec{k}'}\bigr|^2\Bigr\rangle(t)
  \sim \Bigl\langle\,\bigl|\,\phi_{\vec{k}}\bigr|^2\Bigr\rangle(t)
  \sim \frac{T_H}{a(t)\,k^2}. 
\end{align}
When this mode reenters the superhorizon region, 
its amplitude freezes at the value that it takes at $t_k$, 
which is given by 
$T_H/a(t_k)\,k^2\sim m_s^2/k^3$. 
Also, the first factor of the first term in Eq.\ (\ref{separate}), 
$\Bigl\langle\,\bigl|\,\pi_{\vec{k}'}\bigr|^2\Bigr\rangle(t)$, 
keeps the value it had before entering the era of the Hagedorn universe. 
Therefore, the first term of Eq.\ (\ref{separate}) 
is evaluated to be proportional to $k^{-3}$.

For the second term of Eq.\ (\ref{separate}), contrastingly, 
there is a possibility that both of the factors evolve adiabatically. 
A simple evaluation shows that 
the second term is of higher power in $k$ than the first. 
In fact, because $\pi_{\vec{k}'}$ can evolve adiabatically in this case, 
a factor of $a^3(t)$ can appear through the relation 
$\bigl\langle\,\bigl|\,\pi_{\vec{k}'}\bigr|^2\bigr\rangle(t)
\sim a^3(t)\,T_H$, 
which gives an extra factor of $k^3$ factor when evaluated at $t=t_k$.

Collecting all the results obtained above, 
we find that the power spectrum of the velocity field $H a v$ 
behaves as $k^{-3}$ for small $k$. 
Also, the amplitude of the potential $\Psi_{\vec{k}}(t)$ 
can be ignored when evaluated at $t=t_k$, 
because the amplitude of $\Psi_{\vec{k}}$ becomes significant 
only after it enters the superhorizon.

We thus conclude that our model gives  
a scale-invariant power spectrum of the Harrison-Zeldovich type 
for curvature perturbations: 
\begin{align}
 \Bigl\langle\,\bigl|\,\zeta_{\vec{k}}\bigr|^2\Bigr\rangle
  &\sim \Bigl\langle\,\bigl|\,H a v_{\vec{k}}\bigr|^2\Bigr\rangle(t_k)
 \nonumber\\
 &\sim {\rm const.}\,k^{-3}. 
\end{align}

%%%%%%%%%%%%%%%%%%%%%%%%%%%%%%%%%%%%%%%%%%%%%%%%%%%%%%%%%%%%%%%%
%%%%%%%%%%%%%%%%%%%%%%%%%%%%%%%%%%%%%%%%%%%%%%%%%%%%%%%%%%%%%%%%

\resection{Conclusion}

If we trace the evolution of the universe backward in time to the big bang, 
from the point of view of string theory,
it is natural to imagine that the universe undergoes a transition from 
the radiation-dominated era to a period 
characterized by the Hagedorn temperature 
in which many stringy states are excited. 
At the time of this transition, the radius of the universe 
is of the order of $10^{30} l_s$. 
Although this is rather large compared to the string scale, 
the radius of the universe would rapidly shrink to zero
in the Planck time if the evolution of space-time were described by 
the Einstein equation. 
However, the energy density and the curvature during this process 
are so large that the Einstein equation is subject to 
large corrections from stringy effects.
 
Although it is not easy to evaluate the above-mentioned corrections 
rigorously, 
it is natural to hypothesize that in string theory the curvature 
of space-time is bounded by some limiting value. 
For instance, if we take the simplest space-time with a constant curvature, 
i.e. de Sitter space-time, the Hawking temperature is proportional 
to the square root of the curvature, which is simply 
the Hubble parameter of this space-time. 
Then, recalling that temperature is bounded by the Hagedorn temperature, 
we may conclude that curvature of space-time is also bounded. 
If this is true, the shrinking of the universe will not be as rapid as 
that in the naive scenario mentioned above 
but instead, of an exponential form, with the Hubble parameter given 
by the Hagedorn temperature. 
This implies that the behavior of the radius of space-time 
is the same as that of the ordinary inflationary universe, 
while space-time is filled with very dense radiation of 
stringy states. 
This picture is different from the standard picture of inflation. 
In this universe, both the Hawking temperature and the temperature 
of matter fields are equal to the Hagedorn temperature.

Tracing back farther in time, we eventually reach the moment 
at which the radius of the universe becomes of the same order as 
the string scale. 
Then, the universe begins to expand, because in string theory a length 
shorter than the string scale makes no sense. 
This argument is supported by T-duality. 
In this way, we are naturally led to the old idea of 
a cyclic universe. 
However, in our picture, we know the process connecting a big crunch 
and the succeeding big bang, and we can estimate, 
for instance, how much entropy is produced in the combined process of 
one big crunch and one big bang. 
As a result, we arrive at the picture that the universe created 
at the Planck scale has grown into the present size universe 
through the repetition of about forty big crunches and big bangs, 
the number necessary to account for the present amount of entropy.

As was recently pointed out \cite{cyclic2}, 
the idea of a cyclic universe is not at all absurd, 
but indeed a natural conjecture 
when considering certain recent observational results, 
such as the anisotropy of the CMB. 
Although our model is different from that of Ref.\ \cite{cyclic2}, 
it also reproduces the scale-invariant spectrum 
of curvature perturbations 
under the assumption 
that the zero modes of matter fields have long relaxation times.

Other than the anisotropy of the CMB, 
information regarding the early universe has been obtained 
from the baryon number. 
In the Hagedorn universe, because the matter fields have high density 
and are in states very close to thermal equilibrium, 
we can assume that the baryon number is set to zero by the big crunch 
of the previous universe. 
Therefore, it is reasonable to assume that 
in the present universe, the existing baryon number was set at the moment 
when the universe made the transition to the radiation-dominated era, 
right after the Hagedorn universe ended. In fact, as found 
in Ref.\ \cite{Aoki:1997vb}, if the decay of stringy massive particles 
is responsible for the lepton number, we can obtain 
a value of the baryon number consistent with observation.

In our picture, the radius of the universe realizes the minimum length of 
string theory in between a big crunch and big bang. 
There, the sizes of the compactified space-time 
and the four-dimensional space-time 
are of the same order, and they are indistinguishable. 
However, when this state undergoes 
the exponential expansion and the universe enters the radiation-dominated era, 
the radius of the universe is believed to become of the order 
of $10^{30}$ times the string length.
This implies that determination of the space-time dimensionality to be four 
may occur at a certain time close to the moment 
at which the radius of the universe takes the minimum value. 
The reason that the space-time dimension becomes four 
might be because the true vacuum, taking into account nonperturbative 
effects, chooses four-dimensionality. 
However, in opposition to this argument, it may be the case that 
many true vacua exist 
even if nonperturbative effects are incorporated. 
Nevertheless, when our universe underwent the exponential expansion, 
four-dimensionality 
may have been chosen by some yet unknown reason. 
The mechanism responsible for this selection could be either 
related to the maximization of entropy, or due to some dynamical effect, 
such as that discussed by Brandenberger and Vafa \cite{Brandenberger:1988aj}.

In the near future, further developments are expected in observations 
of the very early universe. It may be the right time to contemplate 
what existed before the inflation.

%%%%%%%%%%%%%%%%%%%%%%%%%%%%%%%%%%%%%%%%%%%%%%%%%%%%%%%%%%%%%%%%
%%%%%%%%%%%%%%%%%%%%%%%%%%%%%%%%%%%%%%%%%%%%%%%%%%%%%%%%%%%%%%%%
\section*{\large{Acknowledgment}}

The authors would like to thank Y.~Habara, Y.~Kono, 
A.~Miwa, Y.~Suto and T.~Tada for discussions.

%%%%%%%%%%%%%%%%%%%%%%%%%%%%%%%%%%%%%%%%%%%%%%%%%%%%%%%%%%%%%%%%
%%%%%%%%%%%%%%%%%%%%%%%%%%%%%%%%%%%%%%%%%%%%%%%%%%%%%%%%%%%%%%%%

%%%%%%%%%%%%%%%%%%%%%%%%%%%%%%%
%%% reference
%%%%%%%%%%%%%%%%%%%%%%%%%%%%%%%
\baselineskip=0.7\normalbaselineskip

\end{document}